\documentclass[conference]{IEEEtran}
\IEEEoverridecommandlockouts
\usepackage{booktabs}
\usepackage{graphicx}
\usepackage[utf8]{inputenc}
\usepackage[T1]{fontenc}
\usepackage{multirow}
\usepackage{cite}
\usepackage{amsmath,amssymb,amsfonts}
\usepackage{algorithmic}
\usepackage{graphicx}
\usepackage{textcomp}
\usepackage{xcolor}
\begin{document}

\title{ThermLeT: Transformer-based Temperature Prediction for 2.5D Chiplet Architecture\\

\thanks{}
}
\author{
    \IEEEauthorblockN{
        Varun Darshana Parekh\IEEEauthorrefmark{1}\IEEEauthorrefmark{3}\thanks{\IEEEauthorrefmark{3}Varun Darshana Parekh and Anusha Devulapally contributed equally to this work.},
        Anusha Devulapally\IEEEauthorrefmark{1}\IEEEauthorrefmark{3},
        Sivani Devarapalli\IEEEauthorrefmark{1},
        Cassius Henderson\IEEEauthorrefmark{1},
        Shimeng Yu\IEEEauthorrefmark{2},\\
        Vijaykrishnan Narayanan\IEEEauthorrefmark{1}
    }
    \IEEEauthorblockA{
        \IEEEauthorrefmark{1}The Pennsylvania State University
        \IEEEauthorrefmark{2}Georgia Institute of Technology
    }
}

\maketitle

\begin{abstract}
The drive toward heterogeneous 2.5D chiplet packages promises unmatched performance and customization, yet the resulting density amplifies thermal challenges. To facilitate rapid design space exploration, we present ThermLeT, a multimodal transformer-based, scale invariant surrogate that integrates power density maps, spatial coordinate embeddings, material properties and heat transfer rate to predict steady state temperature in 13 ms, about four orders of magnitude faster than conventional finite element solvers such as ANSYS. To train the transformer model, we also create a dataset of 6,221 chiplet layouts perturbed from three representative architectures. On a 150 sample test set, ThermLeT achieves an RMSE of 3.16°C and an MAE of 2.24°C, improving accuracy by 35\% over ThermEDGe and by 71\% over HotSpot, while running $\approx6\times$ and $11\times$ lower inference latencies respectively. Ablation studies confirm robust generalization to unseen floorplans. Code and dataset available at: https://github.com/MicroSystemsDesignLab/ThermLeT

\end{abstract}


\section{Introduction}

\begin{figure}
\centering
\includegraphics[width=0.8\linewidth]{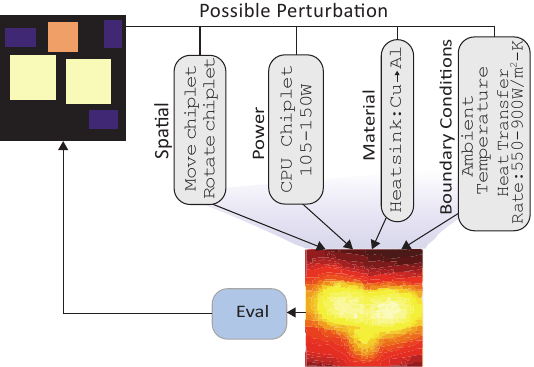}
    \vspace{-0.3cm}
    \caption{Design parameter perturbations in 2.5D chiplet thermal analysis. Four categories of perturbations: spatial, power, material, and boundary conditions require complete thermal recalculation. This iterative evaluation process, while essential for design space exploration, becomes computationally prohibitive when evaluating thousands of configurations using traditional finite element methods.}
    \label{fig:intro}
    \vspace{-0.6cm}
\end{figure}

Chiplet based heterogeneous packages offer modularity, yield, and performance that monolithic systems increasingly struggle to match\cite{sridhar_panchapakesan_system_2022, kumari_ai-enabled_2025}.  
By placing multiple dies on a silicon interposer, 2.5D integration trims interconnect length, cuts energy per bit, and improves signal integrity\cite{refai-ahmed_new_2024}, and recent TSV and multi‑level assembly advances have pushed 2.5D packaging into mainstream production\cite{Shao2018}. \enlargethispage{2\baselineskip} 
Commercial deployments already pair advanced‑node compute chiplets with mature‑node I/O dies (e.g., AMD-EPYC, which combines 7nm CPU cores with a 12nm I/O hub\cite{Naffziger2020}), underscoring the industrial relevance of chiplet design.  
The downside is higher package temperature, because stacked or closely spaced chiplets create local hot spots that degrade reliability and throttle performance\cite{liu_deepoheat_2023}.  Local hot spots accelerate electromigration, timing drift, and leakage, and have become the dominant reliability limiter in modern stacks\cite{liu_deepoheat_2023, Han2022}. 
Achieving optimal thermal performance in 2.5D chiplet systems requires navigating a vast design space defined by spatial, power, material, and cooling parameters, each requiring thermal recalculation to assess feasibility as shown in figure~\ref{fig:intro}. 
Automated floor planners sweep thousands of candidate placements per design cycle\cite{TAP25D, parekh2025stamp25dstructuralthermalaware}. Thermal sign‑off thus depends on prediction tools that can evaluate a new floor plan or power map almost instantly. while traditional approaches constrain designers to limited sampling due to computational costs of 95 seconds per evaluation, machine learning-based thermal prediction enables interactive exploration by reducing evaluation time to milliseconds, fundamentally changing what design optimizations are achievable. 

Recent neural surrogates based on convolutional or graph networks cut latency to a few milliseconds~\cite{9712583, park_hotspot_driven_2024, chhabria_thermal_2021}.  
They remain limited, however, by narrow private datasets and by architectures that struggle with long range heat flow across multiple chiplets and layers.  
A publicly available large‑scale dataset would enable fair comparison and robust training, but none exists today for 2.5D packages.  A single model that serves both interactive floor planning would close a long standing gap in the heterogeneous‑integration (HI) tool chain.  

We answer this need with \emph{ThermLeT}, a Transformer surrogate trained, validated and tested on \textbf{6,221} \textsc{ansys} generated layouts that vary interposer size, chiplet counts, power distributions, material stacks, and cooling conditions.  
Conditioned on power maps, spatial coordinates, and material properties, ThermLeT predicts full chip steady‑state temperature in under \textbf{13\,ms} and lowers root mean square error to \textbf{3.16\,\textdegree C}, while being \textbf{6$\times$} faster than ThermEDGe and \textbf{11$\times$} faster than HotSpot. Although this paper targets temperature prediction, coupling ThermLeT with fast stress surrogates is a promising next step toward full thermo‑mechanical co‑design\cite{parekh2025stamp25dstructuralthermalaware}. Our main contributions are:
\begin{itemize}
    \item A multi modal Transformer model that predicts steady‑state temperature for 2.5D chiplet packages.
    \item \textbf{6,221} \textsc{ansys} derived layouts covering chiplet counts, floor plans, power maps, and material stacks, for reproducible research and benchmarking.
    \item Inference latency under \textbf{13ms}, \textbf{6$\times$} speedup over ThermEDGe, \textbf{11$\times$} over HotSpot, and root mean square error of \textbf{3.16\,\textdegree C} with mean absolute error of \textbf{2.24 \textdegree C} on unseen configurations.
\end{itemize}
The rest of this paper is organised as follows.  
Section~\ref{sec:related} discusses background and reviews related work.  
Section~\ref{sec:method} details the ThermLeT methodology, including dataset generation and model architecture.  
Section~\ref{sec:result} describes the experimental setup, results and analysis.  
Section~\ref{sec:conclude} concludes and provides future direction.

\section{Background and Related Work}
\label{sec:related}
This section first anchors our work in the governing physics of heat conduction and then surveys prior analytical, finite‑element, and learning‑based approaches to thermal prediction for 2.5D chiplet packages.
\begin{table*}[ht]
  \centering
  \caption{Summary of related works with their inputs, approach, and simulation type.}
  \vspace{-0.2cm}
  \label{tab:literature_review}
  \begin{tabular}{c|c|c|c}
    \hline
    \textbf{Related Work} & \textbf{Input} & \textbf{Approach} & \textbf{Steady-State or Transient} \\
    \hline
    ANSYS & Geometry Modeling & FEM/CFD & Both \\
    HotSpot~\cite{hotspot7} & Layout, BC & FEM & Both \\
    DeepOHeat~\cite{liu_deepoheat_2023} & Layout, BC, Power Map & Operator Learning ML & Transient \\
    CoMeT~\cite{siddhu_comet_2022} & Layout, BC, Power Map & FEM & Both \\
    U-Net~\cite{lin_fast_nodate} & Layout, Power Map & Image-to-image and PINN & Steady-State \\
    ThermEDGe~\cite{chhabria_thermal_2021} & Layout, Power Map & Encoder-Decoder Network & Both \\
    POD-ANN~\cite{10909871} & Layout, Power Map, Material Properties & PINN & Steady-State \\
    GCN~\cite{9712583} & Layout, Power Map & Graph Convolutional Networks & Steady-State \\
    \multirow{2}{*}{\textbf{ThermLeT (Ours)}} 
      & Layout, Power Map, Material Properties 
      & \multirow{2}{*}{Transformer-based Network} 
      & \multirow{2}{*}{Steady‐State} \\
      & Coordinates, Heat Transfer Rate 
      &  &  \\
    \hline
  \end{tabular}
  \vspace{-0.5cm}
\end{table*}
\subsection{Heat Conduction Formulation}
The temperature field addressed in this work satisfies the three dimensional heat conduction equation: $ \nabla \!\cdot\! \bigl(k\nabla T\bigr) + q = 0 $, 
where $k$ is the thermal conductivity and $q$ is the volumetric heat source density.  
Chiplet and interposer surfaces that face the heatsink or ambient air follow a convection boundary condition with $T_{\infty}$ as ambient temperature: $-k\nabla T\!\cdot\!\mathbf{n} = h\,(T - T_{\infty}),$
while the remaining lateral and bottom faces are treated as adiabatic. The left hand side of equation represents normal heat flux density leaving the solid  with n representing outward normal unit vector to the surface and right hand side represent convective heat flux absorbed by the environment with h convection heat transfer coefficient (W/m²·K).
Solving this partial differential equation with discontinuous material properties needs fine meshes and iterative finite element solvers.

\subsection{Literature Review}
Thermal modeling for 2.5D chiplet packages has progressed through several distinct generations. First, high-fidelity finite-element tool like ANSYS remain the gold standard for accuracy but incur minutes per design, spurring development of near real-time surrogates. Table~\ref{tab:literature_review} outlines common models found in the literature.\\
HotSpot\cite{hotspot7} builds a lumped thermal resistance network modeling each block as an isothermal node. It supports both steady and transient modes and runs in milliseconds on a CPU, but tends to produce step-like artifacts and significant error under fine power patterns. To accelerate temperature prediction, physics-informed operator-learning frameworks like DeepOHeat~\cite{liu_deepoheat_2023} leverage neural operators trained on FEA data. However they require complex domain decompositions and equation-based models can compromise on accuracy to reproject the data into a form without discontinuities\cite{sultan_survey_2019}.\\
Convolutional encoder–decoder networks such as ThermEDGe~\cite{chhabria_thermal_2021} reduce inference to milliseconds by mapping fixed-resolution power and layout images directly to temperature fields. However, because it assumes a fixed millimetre‑to‑pixel resolution, the model must be retrained when the interposer footprint changes.  Graph-based approaches like  Dual-Channel MPNNs~\cite{li_transient_2024} and steady-state GCNs~\cite{9712583} improve on irregular geometries and capture local adjacency, yet they lack a principled way to integrate boundary conditions and exhibit high variance in runtime as die size scales.  \\
More recently, multimodal Transformers have demonstrated state-of-the-art fusion of heterogeneous inputs~\cite{xu_multimodal_2023}, but to date they have not been adapted for chiplet thermal analysis.  Comprehensive surveys~\cite{chen-survey} also underscore the need for open benchmarks and co-optimization frameworks spanning thermal, electrical, and mechanical co-design in heterogeneous integration.  \\
In this paper, we bridge these gaps by introducing a unified, end-to-end differentiable Transformer surrogate that jointly attends to dense power density maps, explicit 2D coordinate embeddings, and scalar material and cooling parameters.  Our architecture delivers accurate steady-state temperature predictions across arbitrary chiplet layouts without any footprint-specific retraining, while simultaneously preserving fine-grained spatial detail that prior surrogates cannot capture in a single framework.  

\section{Methodology}
\label{sec:method}
We describe the complete ThermLeT pipeline, covering ground‑truth dataset generation, input pre‑processing, and the proposed multimodal Transformer architecture.
\subsection{Dataset Generation \& Pre-processing}
\label{sec:data}
\begin{table}[t]
\centering
\caption{Material Properties for Key Components}
\label{tab:material_properties}
\vspace{-0.3cm}
    \renewcommand{\arraystretch}{1.2}
    \setlength{\tabcolsep}{2pt} 
\begin{tabular}{cccccc}
\hline                                  \textbf{Components} & \textbf{Materials} & \begin{tabular}[c]{@{}l@{}}\textbf{Thermal}\\ \textbf{Conductivity}\\ \textbf{(w/m.k)}\end{tabular} & \begin{tabular}[c]{@{}l@{}}\textbf{CTE}\\ \textbf{(ppm/°C)}\end{tabular} & \begin{tabular}[c]{@{}l@{}}\textbf{Young's }\\ \textbf{Modulus}\\ \textbf{(gpa)}\end{tabular} \\\hline
\textbf{Substrate}  & FR-4                     & 0.3                    & 13.0                   & 20 \\ 
\textbf{C4 Bump}   & Tin-Lead (60-40)          & 50                     & 20.5                   & 20 \\ 
\textbf{$\mu$ Bump}  & SAC (SnAgCu)              & 50                     & 20                      & 50  \\ 
\textbf{Interposer} & Silicon Anisotropic       & 148                    & 2.6                     & 150\\ 
\textbf{Chiplets}   & Silicon Anisotropic       & 150                   & 3.1                     & 130 \\ 
\textbf{TIM}        & Indium                    & 86                     & 29                      & 10 \\ 
\textbf{Heat Sink}  & Copper Alloy             & 398                   & 16                      & 120\\ 
\textbf{Underfill}  & SiO2                      & 1.4                    & 0.5                     & 80 \\ \hline
\end{tabular}
\vspace{-0.2cm}
\end{table}

\begin{figure}
    \includegraphics[width=1.05\linewidth]{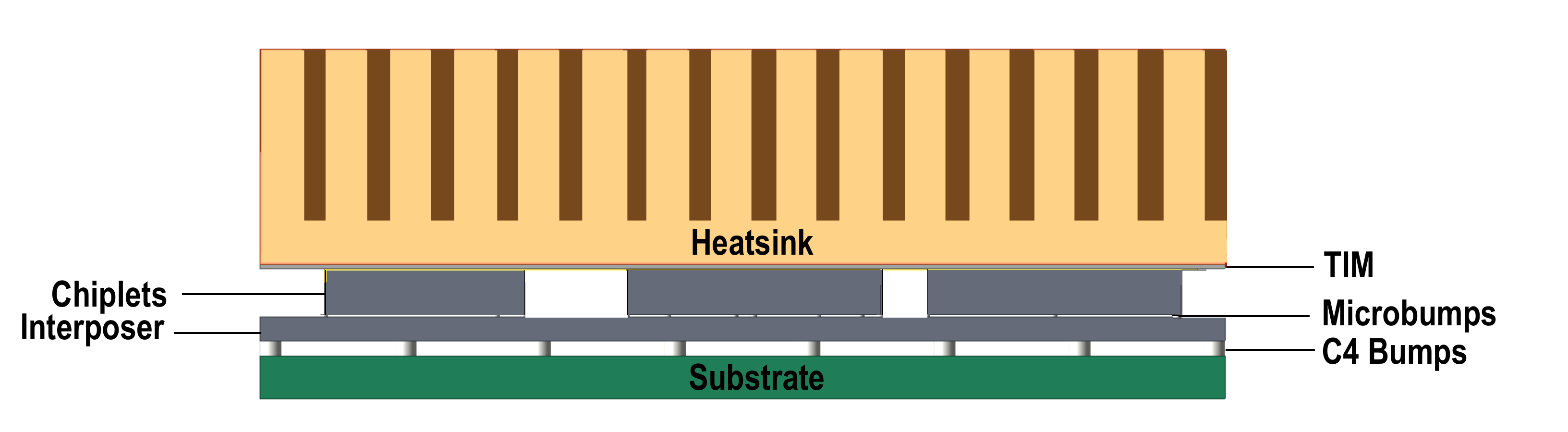}
    \vspace{-0.75cm}
    \caption{Cross-sectional schematic of the 2.5D chiplet integration stack. The structure includes key components such as C4 bumps, thermal interface materials (TIM), underfill, interposer, microbumps, chiplets, and a heatsink. This model serves as the foundation for finite element analysis (FEA) simulations}
    \label{fig:modelling}
    \vspace{-0.6cm}
\end{figure}
A supervised corpus of 6221 chiplet placements were built for training and evaluating ThermLeT. Three representative 2.5D architectures are used - \textit{Micro150, Ascend910, and MultiGPU}. The dimension and power details of the architecture discussed in table III. For dataset generation, the chiplet center positions and orientations are perturbed.
For each perturbed layout, its geometry is generated and exported to ANSYS Mechanical, which solves for its steady-state temperature on a package with a silicon interposer (50 mm × 50 mm, 45 mm × 45 mm for Ascend910, 100 \textmu m thickness), Indium based Thermal Interface Material (TIM) and copper heatSink under forced convection (580/720/950 $W/m^{2}K$) at a 23$^\circ$C ambient. The resulting thermal and stress outputs were interpolated onto a uniform 224 × 224 grid to form the target tensor. The modeling and material details are tabulated in figure~\ref{fig:modelling} and table~\ref{tab:material_properties} respectively.
\begin{table}[t]
  \centering
  \caption{Chiplet dimensions and powers in 2.5D examples.}
  \vspace{-0.2cm}
  \setlength{\tabcolsep}{3.3pt} 
  \label{tab:chiplet-dimensions}
  \begin{tabular}{@{} l ccc cc ccc @{}}
    \toprule
    & \multicolumn{3}{c}{Multi-GPU}
    & \multicolumn{2}{c}{Micro150}
    & \multicolumn{3}{c}{Ascend 910} \\
    \cmidrule(lr){2-4} \cmidrule(lr){5-6} \cmidrule(lr){7-9}
    Chiplet      & CPU  & GPU   & HBM   & CPU   & DRAM  & Virtuian & Nimbus & HBM   \\
    \midrule
    Widths [mm]  & 12   & 18.2  & 7.75  & 8.25  & 8.75  & 31.4     & 10.5   & 7.75  \\
    Height [mm]  & 12   & 18.2  & 11.87 & 9     & 8.75  & 14.5     & 16     & 11.87 \\
    Power [W]    & 105  & 295   & 20    & 150   & 20    & 256      & 14     & 20    \\
    \bottomrule
  \end{tabular}
  \vspace{-0.55cm}
\end{table}
For each sample, the input tensor comprises five feature maps on the same grid: power density, material thermal conductivity, effective heat-transfer coefficient, and normalized x \& y coordinate maps. Thus, each input–target pair has dimensions 224 × 224 × 5 and 224 × 224 × 3, respectively. Importantly, this methodology is inherently scalable and is not tied to any fixed interposer size. By encoding spatial information through multimodal inputs, including normalized coordinate maps and pixel-to-physical scaling (pixels/mm), ThermLeT learns spatial correlations that generalize across varying interposer geometries and resolutions.
\vspace{-0.1cm}
\subsection{Model Architecture}
\begin{figure}
    \centering
\includegraphics[width=0.80\linewidth]{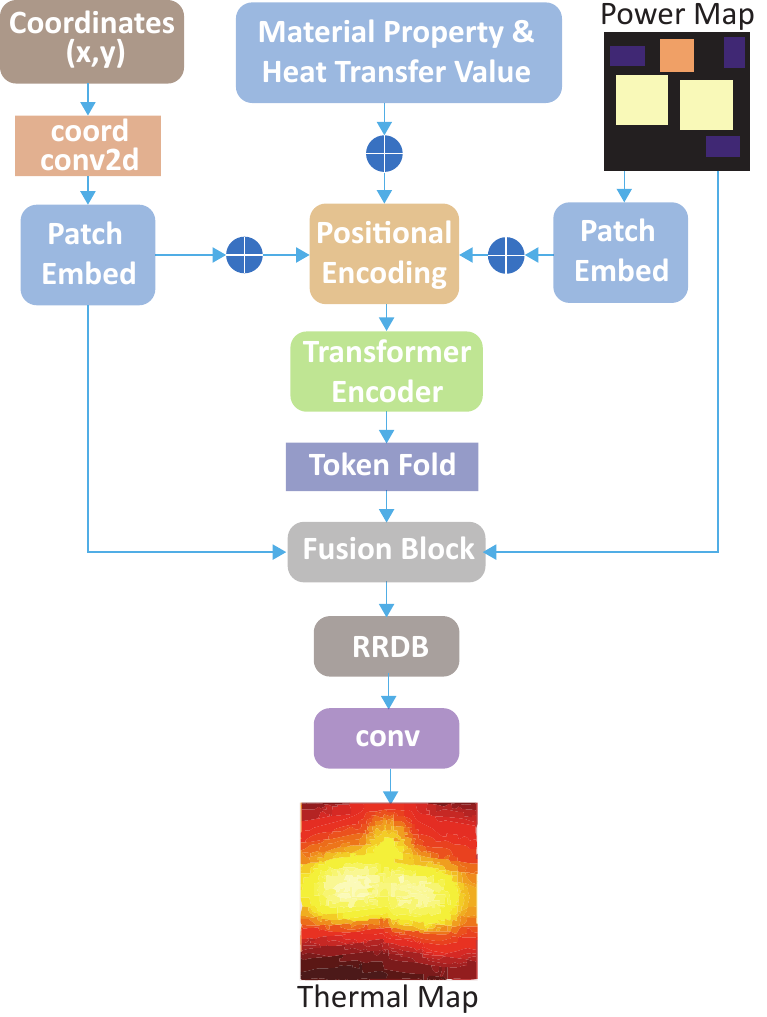}
\vspace{-0.3cm}
    \caption{Overview of the proposed ThermLeT architecture. The model takes three input modalities: power map, spatial coordinates, and material properties and embeds them using dedicated modules. These embeddings are fused and passed through a Transformer encoder and convolutional decoding blocks to predict the final temperature map. The design enables the model to capture spatial and physical dependencies for accurate thermal prediction across diverse chiplet architectures.}
    \label{fig:network_architecture}
    \vspace{-0.65cm}
\end{figure}
Early thermal surrogates, from lumped‐resistance networks like HotSpot to convolutional encoder–decoders and graph‐based MPNNs, have shaved inference time to milliseconds but still lack the receptive field to capture long‐range heat diffusion across multiple chiplets and layers, struggle to fuse dense spatial inputs (power maps, coordinates) with sparse global scalars (material and boundary properties), and generalize poorly to unseen floorplans or material stacks. In contrast, Transformers excel at modeling long‐range dependencies and multimodal fusion~\cite{Devulapally_2024_CVPR}, motivating our adaptation of this prior framework to the thermal‐prediction problem.\\
Figure~\ref{fig:network_architecture} illustrates the resulting pipeline. We begin by co‐registering three aligned inputs on an $H\times W$ grid: a power density map $P\in\mathbb{R}^{H\times W}$ encoding per‐pixel heat dissipation, a 2D coordinate map $C\in\mathbb{R}^{H\times W\times 2}$ whose channels carry normalized $(x,y)$ positions and a pair of scalar architecture features: thermal conductivity $k$ and interface heat‐transfer coefficient $h$, that globally define material and cooling conditions. To instill absolute positional awareness, the raw coordinate map is first enriched via a \texttt{CoordConv2d} layer which concatenates learned coordinate channels to the inputs.\\
Both the augmented coordinate map and the power density map are partitioned into non-overlapping patches and projected through separate \texttt{PatchEmbed} modules into a shared \(D\)-dimensional token space.  In parallel, the scalar pair \((k,h)\) is embedded via a small feed-forward network into the same \(D\)-dimensional latent space. Finally, learnable positional encodings are added to the concatenated token sequence to preserve the underlying 2D layout.\\
The fused sequence of $N$ tokens is processed by a stack of six ViT‐style encoder blocks. Each block employs 12‐head multi‐head self‐attention, a two‐layer feed‐forward subnetwork with GELU activations, residual connections, and layer normalization. This design enables each patch token to attend jointly to local power, spatial coordinates, and global material parameters.
\\
After encoding, we fold the enriched tokens back into $H\times W$ feature maps via an inverse patch‐fold and merge these maps with skip‐connected convolutions of the original inputs, modality‐specific fusion blocks align power‐derived and coordinate‐derived features, an RRDB refines them, and a final two‐layer convolutional head with ReLU produces the predicted steady‐state temperature field.\\
By building directly on our prior multimodal Transformer work~\cite{Devulapally_2024_CVPR}, this end‐to‐end architecture not only overcomes the limitations of earlier surrogates but also remains highly modular-enabling future extensions such as stress or mechanical inputs while achieving state‐of‐the‐art accuracy.
The network is trained using a combination of two loss functions. The primary term is the Smooth L1 loss (Huber loss), applied over valid pixels: 
\vspace{-0.2cm}
\begin{equation}
L_{\mathrm{l1\_loss}} = \sum_{i=1}^{N} \left| D_k(i) - \hat{D_k}(i) \right|
\end{equation}
where $D_k(i)$ and $\hat{D_k}(i)$ denote the ground truth and predicted temperatures, and $N$ is the number of valid pixels.\\
To enforce structural consistency, a gradient matching loss term is added and computed at multiple spatial scales: 
\vspace{-0.1cm}
\begin{equation}
L_{\text{grad}} = \frac{1}{N} \sum_{s} \sum_{i} \left| \nabla_x R_k^s(i) \right| + \left| \nabla_y R_k^s(i) \right|
\end{equation}
where $R_k = D_k - \hat{D_k}$ is the residual map, and $\nabla_x$, $\nabla_y$ denote spatial gradients along $x$ and $y$ computed using Sobel operators over four resolution levels.\\
The final training objective combines these losses as:
\vspace{-0.1cm}
\begin{equation}
L_{\mathrm{total}} = \lambda \cdot L_{\mathrm{l1\_loss}} + \beta \cdot L_{\mathrm{grad}}
\end{equation}
where $\lambda = 0.5$ and $\beta = 0.25$ are weighting coefficients. 
\section{Result and Analysis}
Finally, we evaluate ThermLeT against state‑of‑the‑art baselines on three chiplet architectures, detailing the experimental setup, quantitative results, ablation studies, and performance analysis.
\label{sec:result}

\subsection{Experimental Setup}
\label{sec:experimental-setup}

 The experiments are conducted on a dataset comprising 6221 samples generated from three different architecture configurations. Out of the total, 150 samples are reserved for testing and the remaining 6071 samples are randomly split into 80\% for training and 20\% for validation. Normalization is not applied to the data to maintain the original scale of the temperature and power inputs throughout the training and evaluation process. All Experiments are conducted on a single NVIDIA A100 GPU.

 The proposed model, ThermLeT, is implemented in PyTorch and trained using the Adam optimizer with a learning rate of 0.0003, weight decay of 1e-4, and batch size of 16 for 30 epochs. Data augmentation includes random horizontal and vertical flips (with 0.5 probability) and Gaussian noise on power maps. A gradient-based smoothness loss is added to promote spatial consistency in predicted temperature maps.
 
 The model is evaluated on using two standard regression metrics: Mean Absolute Error (MAE) and Root Mean Squared Error (RMSE), computed over $N$ valid pixels between the predicted temperature $\hat{T}(i)$ and ground truth $T(i)$. The corresponding formulas are: $\text{MAE} = \frac{1}{N} \sum_{i=1}^{N} \left| T(i) - \hat{T}(i) \right|$ and $\text{RMSE} = \sqrt{\frac{1}{N} \sum_{i=1}^{N} \left( T(i) - \hat{T}(i) \right)^2} $
Along with the regression metrics, the performance metrics were also calculated such as latency, throughput and FLOPs. 

\subsection{Baselines}
We compare our models with two baselines: ThermEDGe~\cite{chhabria_thermal_2021} and HotSpot7.0~\cite{hotspot7}. We perform a detailed comparative evaluation with ground truth FEM simulations using ANSYS. To ensure fair comparison, all methodologies were carefully aligned in terms of thermal boundary conditions and material properties. Specifically, parameters such as the convection heat transfer coefficient ($h$), ambient temperature ($T_{\infty}$), and material layer resistances were kept consistent across methods.
\subsubsection{HotSpot} 
A lumped vertical thermal resistance model of 32 x 32 grid resolution is used by explicitly computing the convection thermal resistance ($R_{\text{convec}}$) as $R_{\text{convec}} = 1/(hA_{\text{wetted}})$, where $A_{\text{wetted}}$ represents the actual wetted surface area of the heatsink, excluding the interface with the TIM. From the ANSYS model, the wetted surface areas were precisely calculated, and the HotSpot configuration was closely monitored. For instance, with a heat transfer coefficient of $950\text{W/m}^2\text{K}$ and a heatsink wetted area of $14,361\,\text{mm}^2$, we obtained a convection resistance of $0.074\text{K/W}$.
\subsubsection{ThermEDGe}
We re‐implemented ThermEDGe [21], a CNN‐based encoder-decoder for steady-state temperature prediction. To enable a direct comparison, we updated the original 32×32‐input code to accept 224×224 power maps. This baseline model processes only the power map and was trained under the default settings for 50 epochs.
\subsection{Results}
\begin{figure*}
    \centering    \includegraphics[width=0.8\linewidth]{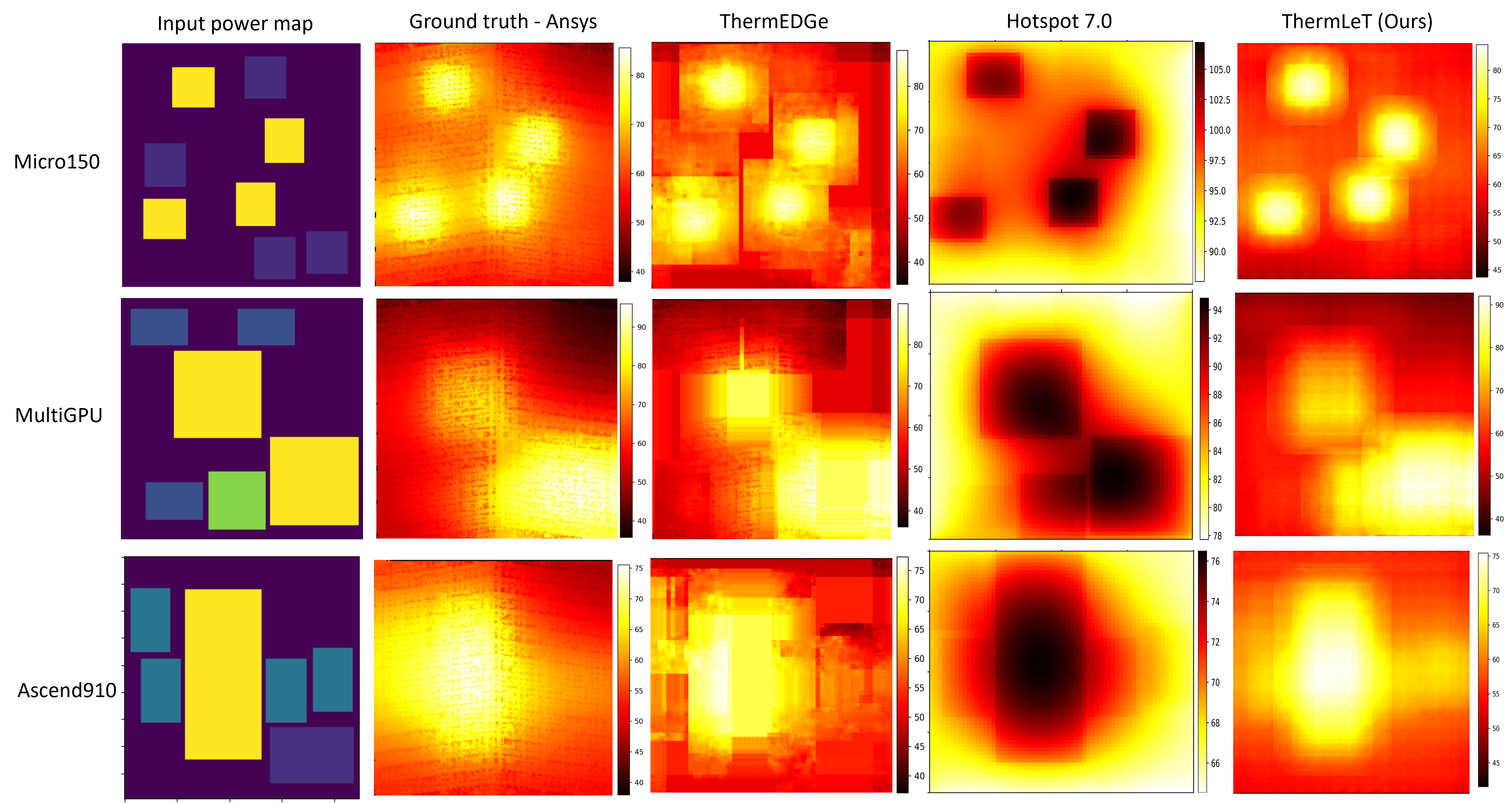}
    \vspace{-0.2cm}
    \caption{Visual comaprision of temperature predictions: (a) Input Power map with different color representing different power values shown ~\ref{tab:chiplet-dimensions} (b) Ground truth generated using Ansys simulation (c,d) Temperature predictions by baseline ThermEDGe\cite{chhabria_thermal_2021} and Hotspot \cite{hotspot7} (e) Proposed transformer based temperature predictor.}
    \label{fig:visual_results}
    \vspace{-0.7cm}
\end{figure*}
We evaluate our approach on a test set of 150 samples, encompassing inputs from three chiplet architectures: Micro150, Ascend910, and MultiGPU.
\begin{table}[ht]
\vspace{-0.2cm}
\centering
  \caption{Evaluation metrics on the test set of 150 samples. Lower values indicate better performance.}
  \vspace{-0.2cm}
  \label{tab:chiplet_eval}
  \begin{tabular}{c|c|c}
    \hline
    Method & RMSE ↓ & MAE ↓ \\
    \hline
    ThermEDGe~\cite{chhabria_thermal_2021} & 4.893 & 3.691 \\
    HotSpot~\cite{hotspot7} & 10.734 & 9.886\\
    ThermLeT (Power Only) & \underline{4.1186} & \underline{3.307} \\
    ThermLeT (Ours)& \textbf{3.160} & \textbf{2.238}\\
    \hline
  \end{tabular}
  \vspace{-0.3 cm}
\end{table}
%
Table~\ref{tab:chiplet_eval} reports RMSE and MAE for HotSpot, the ThermEDGe baseline, and our two ThermLeT variants. HotSpot performs worst with an RMSE 10.734°C, MAE 9.886°C for peak temperature, while ThermEDGe reduces errors to RMSE 4.893°C and MAE 3.691°C. Our power-only ThermLeT variant further lowers error to RMSE 4.119 °C, which is 16\% lower than ThermEDGe and 62\% lower than HotSpot and MAE 3.307 °C, which is 10\% lower than ThermEDGe and 67\% lower than HotSpot. Finally, the full multi-modal ThermLeT achieves the best accuracy of RMSE 3.160 °C, MAE 2.238 °C, representing 35\% and 39\% reductions versus ThermEDGe, and 71\% and 77\% reductions versus HotSpot.

Figure 3 presents three chiplet layouts, Micro150 (top), MultiGPU (middle), and Ascend910 (bottom) compared with their input power maps, ANSYS ground-truth temperature fields, and predictions from ThermEDGe, HotSpot 7.0, and our transformer-based ThermLeT. The ANSYS simulations reveal smoothly varying thermal gradients around high-power blocks and sharply defined hotspots that correspond to elevated power densities. ThermEDGe locates the correct hotspot regions but introduces blocky artifacts and underestimates peak temperatures due to its coarse low-resolution routing of power densities. HotSpot 7.0 also applies a compact block-based thermal RC network in which each floorplan block is modeled as an isothermal node with uniform power density. This causes abrupt temperature jumps at block boundaries and sharp, step-like gradients. In contrast, ThermLeT aligns almost pixel-to-pixel with the ANSYS ground truth. It preserves crisp hotspot shapes, recovers maximum temperatures with high accuracy, and captures subtle spatial–thermal nuances even in densely packed GPU regions. This performance demonstrates ThermLeT’s ability to learn fine-grained multimodal power, material and layout inputs.
\begin{table}[ht]
  \centering
  \vspace{-0.2cm}
  \scriptsize
  \caption{Latency, throughput, and FLOPs for ThermLeT and baselines on the full dataset.}
  \vspace{-0.2cm}
  \label{tab:chiplet_perf}
  \resizebox{0.8\linewidth}{!}{%
    \begin{tabular}{@{}cccc@{}}
      \toprule
      \textbf{Model} & \textbf{Latency} & \textbf{Throughput} & \textbf{FLOPs} \\
                     & (s/sample) & (samples/s) & (GFLOPs) \\
      \midrule
      Ansys steady-state thermal & 95.000* & - &\\
      ThermEDGe~\cite{chhabria_thermal_2021} & 0.0762  & 13.11  & \textbf{11.42} \\
      HotSpot~\cite{hotspot7}              &   0.1442*    &   -    &   -   \\
      ThermLeT (Ours)                         & \textbf{0.0130}  & \textbf{76.77 } & 155.79 \\
      \bottomrule
    \end{tabular}%
  }
  \\ 
  \vspace{0.1cm}\text{*}HotSpot and Ansys is run on Intel 14 core i9-1090X CPU and 11th Gen Intel Core i7-11700 4 core respectively. 
  \vspace{-0.5cm}
\end{table}
Table~\ref{tab:chiplet_perf} summarizes inference efficiency on the held-out test dataset. Ansys’s steady-state FEM solver requires 95 000 ms per sample, highlighting the impracticality for large-scale deployment. Among learned models, ThermEDGe processes each test sample in 76.27 ms per sample with a 11.42 GFLOPs cost, while HotSpot takes 144.20 ms per sample. Our full multi-modal ThermLeT reduces latency to 13.03 ms per sample, a 5.9× speed-up over ThermEDGe and an 11.1× speed-up over HotSpot-boosting throughput to 76.77 samples/s. This runtime gain comes at a computational cost of 155.79 GFLOPs per sample, approximately 13.6× the FLOPs of ThermEDGe.

\subsection{Ablation Study}
Two ablation studies are conducted to evaluate the design choices and effectiveness of the proposed model, ThermLeT.
\subsubsection{Varying Input Modalities}
To assess the impact of each input modality, we trained two ablated ThermLeT variants, power + coordinates and power + material properties, and compared them against the full three-modal model (Table~\ref{tab:varying_input_mod}). Adding material properties to power yields about a 12\% error reduction over the power + coordinates variant, indicating that physical context (thermal conductivity and convection) is more informative on its own than raw spatial cues. Crucially, fusing all three inputs drives a further 42\% RMSE and 48\% MAE reduction compared to the stronger two-modal setup. 

\begin{table}[ht]
\vspace{-0.3cm}
  \centering
  \caption{RMSE and MAE for different input combinations. The full model uses power, coordinates, and material properties.}
  \vspace{-0.2cm}
  \label{tab:varying_input_mod}
  \resizebox{0.85\linewidth}{!}{%
    \begin{tabular}{@{}lcc@{}}
      \toprule
      \textbf{Input Modalities} & \textbf{RMSE ↓} & \textbf{MAE ↓} \\
      \midrule
      Power + Coord + MatProp (Full Model) & \textbf{3.160} & \textbf{2.238} \\
      Power + MatProp                      & 5.424          & 4.271 \\
      Power + Coord                        & 6.162          & 4.858 \\
      \bottomrule
    \end{tabular}%
  }
  \vspace{-0.4cm}
\end{table}
\subsubsection{Generalization Across Chiplet Architectures}
A thermal‐prediction model must generalize to unseen chiplet layouts-differing in interposer size, floorplan geometry, power distribution without retraining. To assess the generalizable the models are to unseen chiplet configurations was conducted. To simulate this real‐world scenario, we therefore conduct an ablation in which both ThermEDGe and ThermLeT are trained exclusively on MultiGPU samples (50×50 interposer) and then tested on the unseen Ascend910 (45×45 interposer) and Micro150 configurations. Table~\ref{tab:chiplet_generalization} shows that ThermLeT outperforms the baseline in both RMSE and MAE, demonstrating stronger robustness to architectural shifts. These results highlight ThermLeT’s ability to learn transferable, physically grounded representations that generalize well across heterogeneous hardware platforms.
\begin{table}[ht]
\vspace{-0.3cm}
  \centering
  \caption{RMSE and MAE on unseen chiplet architectures (Ascend910 + Micro150) after training only on MultiGPU.}
  \vspace{-0.2cm}
  \label{tab:chiplet_generalization}
  \begin{tabular}{ccc}
    \hline
    Method & RMSE ↓ & MAE ↓ \\
    \hline
ThermEDGe~\cite{chhabria_thermal_2021} & 9.6493 & 7.6973 \\
    ThermLeT (Ours)                         & \textbf{8.908} & \textbf{7.3745} \\
    \hline
  \end{tabular}
  \vspace{-0.6cm}
\end{table}
\section{Conclusion}
\label{sec:conclude}
ThermLeT delivers a paradigm shift in rapid thermal evaluation for 2.5D chiplet architectures: by fusing power maps, spatial coordinates, and material properties, it achieves sub-13 ms inference with 76.77 samples/s while maintaining accuracy (RMSE 3.16$^\circ\text{C}$, MAE 2.24$^\circ\text{C}$). This performance not only outpaces ThermEDGe and HotSpot by over 5× in latency and 11× in accuracy, but also generalizes seamlessly to unseen chiplet layouts. In future work, we will enhance ThermLeT with physics-aware transformer modules that inject explicit heat-transfer constraints into the attention mechanisms, and extend the architecture with a dedicated stress-estimation head to predict thermomechanical stresses alongside temperature fields for a unified AI-driven thermal-mechanical co-analysis tool.
\vspace{-0.4cm}
    
\section{Acknowledgement}
This work was supported by PRISM,  one of the seven centers in JUMP 2.0, a Semiconductor Research Corporation (SRC) program sponsored by DARPA.

\bibliographystyle{IEEEtran}
\bibliography{ref.bib}
\end{document}